\documentclass[10pt,letterpaper]{article}
\usepackage{ccn}
\usepackage{pslatex}
\usepackage{apacite}
\usepackage{graphicx}
\usepackage{enumitem}   
\graphicspath{ {./images/} }
\begin{document}
\title{Reconstructing seen images from human brain activity via guided stochastic search}
 
\author{{\large \bf Reese Kneeland (rek@umn.edu)} \\
  Department of Computer Science, University of Minnesota\\
  Minneapolis, MN 55455 USA
  \AND {\large \bf Jordyn Ojeda (ojeda040@umn.edu)} \\
  Department of Computer Science, University of Minnesota\\
  Minneapolis, MN 55455 USA
\AND {\large \bf Ghislain St-Yves (gstyves@umn.edu)} \\
  Department of Neuroscience, University of Minnesota\\
  Minneapolis, MN 55455 USA
\AND {\large \bf Thomas Naselaris (nase0005@umn.edu)} \\
  Department of Neuroscience, University of Minnesota\\
  Minneapolis, MN 55455 USA}

\maketitle
\pagebreak

\clearpage

\section{Abstract}
{
\bf
Visual reconstruction algorithms are an interpretive tool that map brain activity to pixels. Past reconstruction algorithms employed brute-force search through a massive library to select candidate images that, when passed through an encoding model, accurately predict brain activity. Here, we use conditional generative diffusion models to extend and improve this search-based strategy. We decode a semantic descriptor from human brain activity (7T fMRI) in voxels across most of visual cortex, then use a diffusion model to sample a small library of images conditioned on this descriptor. We pass each sample through an encoding model, select the images that best predict brain activity, and then use these images to seed another library. We show that this process converges on high-quality reconstructions by refining low-level image details while preserving semantic content across iterations. Interestingly, the time-to-convergence differs systematically across visual cortex, suggesting a succinct new way to measure the diversity of representations across visual brain areas.

}
\begin{quote}
\small
\textbf{Keywords:}
decoding; vision; generative models; diffusion models; fMRI
\end{quote}

\section{Introduction}

Successfully reconstructing even a moderately complex image from its evoked brain activity requires a strong and appropriate prior. In \cite{NASELARIS2009902}, we used a massive image library as the prior. We used an encoding model for early areas to screen the library for images with low-level details that were consistent with brain activity in early visual areas. A high-level encoding model was then used to select images that were also consistent with brain activity in higher visual areas. This “structure before semantics” strategy was an inversion of reconstruction priorities, as it produced semantically correct reconstructions only in cases where low-level details and semantic content were highly correlated in the image library. This made the success of the method dependent on the content of the library.

We use a recently developed generative model \cite{stablediffusion} to impose a strong prior that guarantees naturalistic reconstructions with interpretable content. We further take advantage of this model to induce a “semantics before structure” search. We search through an image library generated by the diffusion model with guidance from a semantic CLIP embedding \cite{radford2021learning} decoded from brain activity. We then use encoding models for early visual areas\cite{St-Yves_heirarchy} to iteratively refine the library until convergence on a final reconstruction.

\begin{figure}[!htb]
\begin{center}
\includegraphics[scale=0.18]{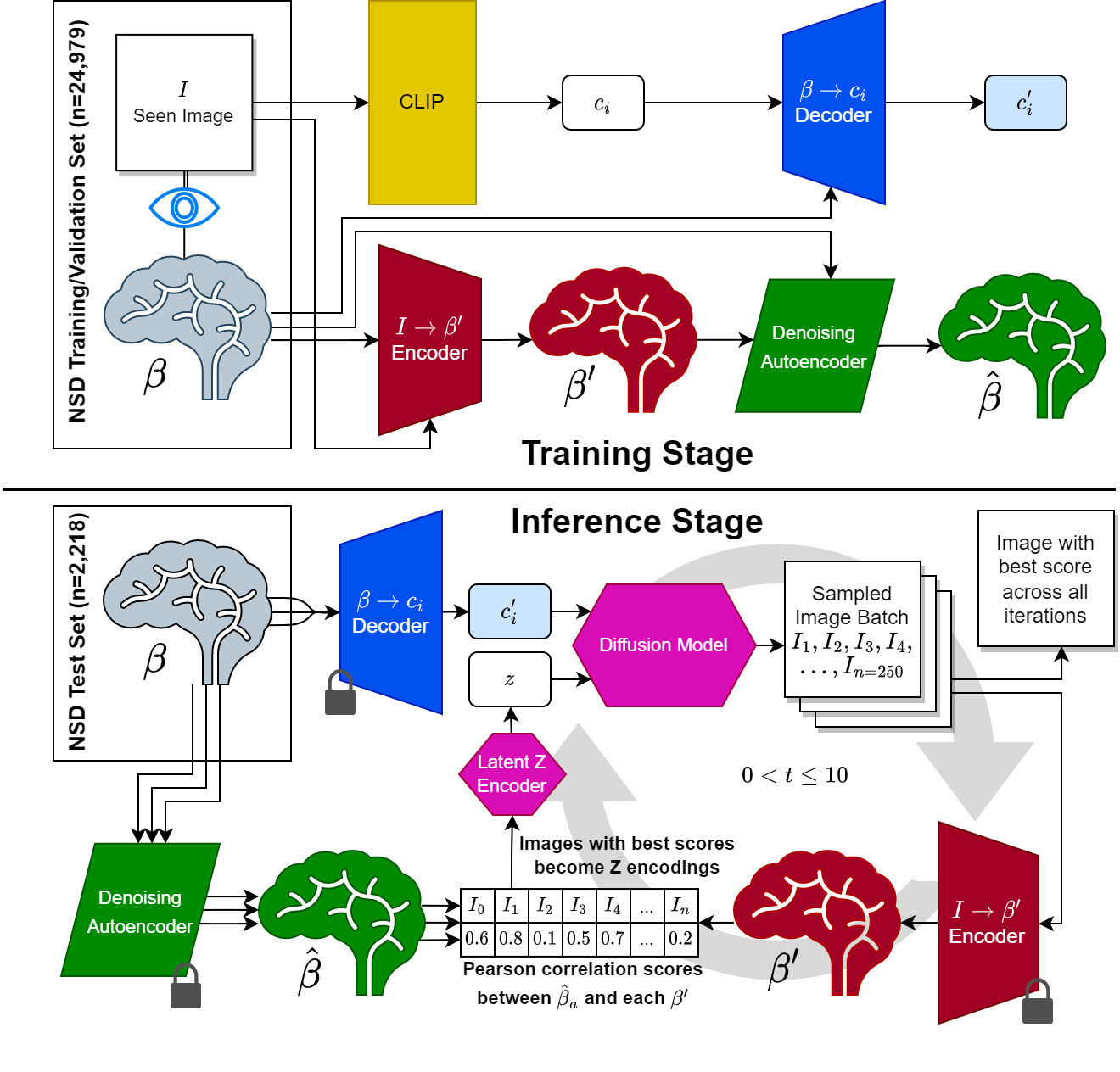}
\end{center}
\caption{Pipeline diagram for our method: the top half demonstrates the Training stage for our models, and the bottom half depicts the Inference Stage that deploys our stochastic search procedure.} 
\label{figure:results}
\end{figure}

\begin{figure}[!ht]
\begin{center}
\includegraphics[scale=0.07]{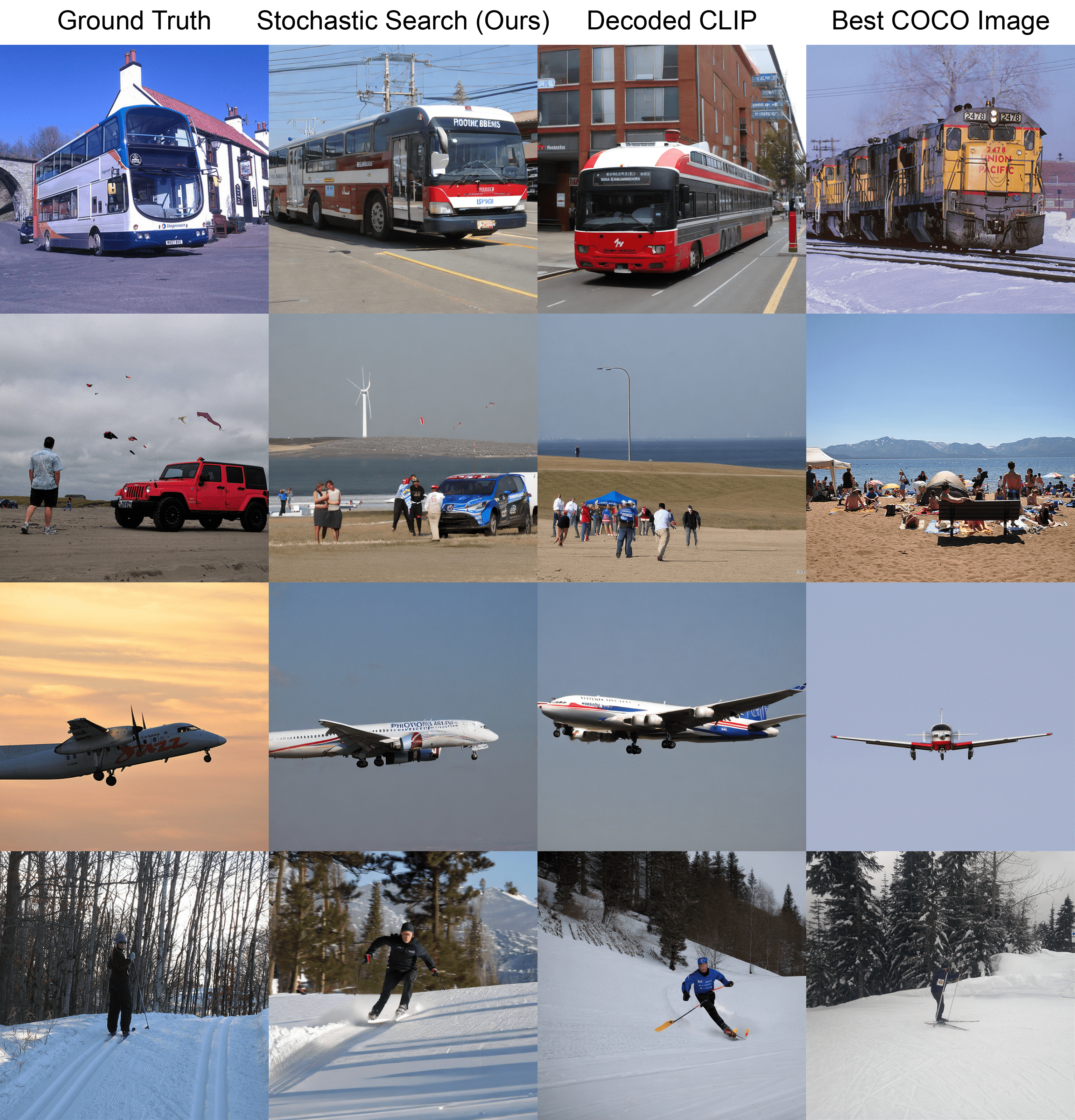}
\end{center}
\caption{Qualitative comparison of our method against a CLIP decoding-only method, and brute-force search through a library of COCO images. \cite{microsoftcoco}} 
\label{sample-figure}
\end{figure}

\begin{figure*}[!ht]
\begin{center}
\includegraphics[scale=0.047]{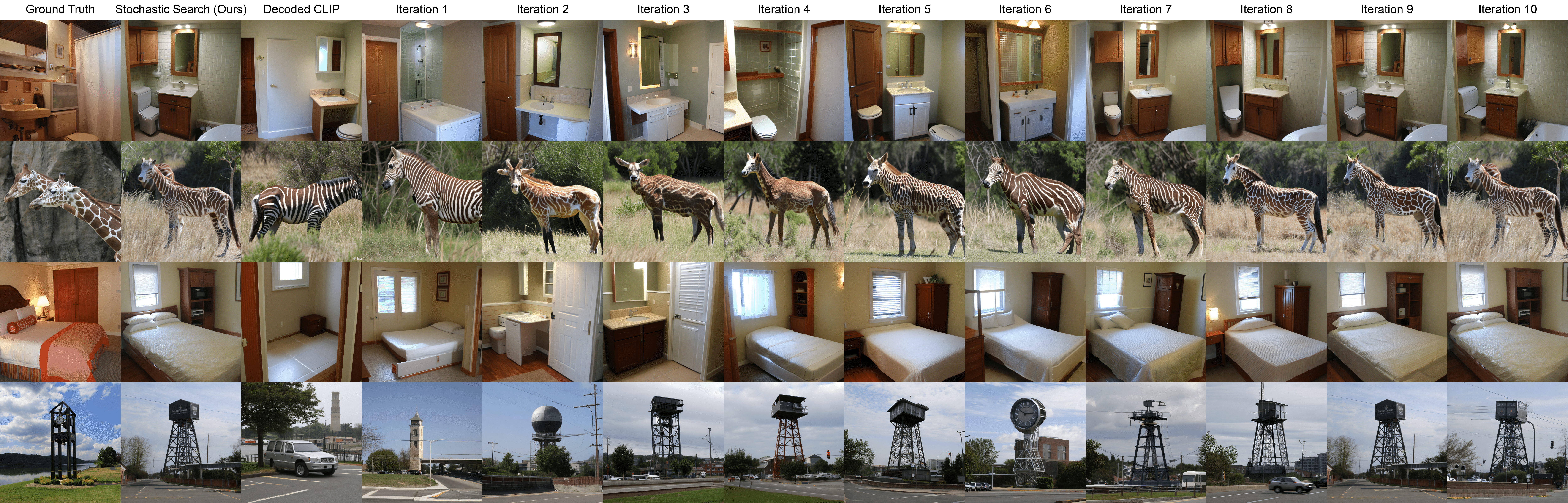}
\end{center}
\caption{Refinement of reconstructions over search iterations (left to right) } 
\label{figure:iterations}
\end{figure*}

\section{Methods}
We analyze the Natural Scenes Dataset \cite{Allen2021a}. For the subject shown here (S1), we partition the data into training (n=20,809, used for encoding/decoding model parameter estimation), validation (n=4,170, used for model regularization), hyperparameter (n=552, used to fine-tune the iterative search procedure), and test (n=2,218, used to assess decoding performance) sets. Hyperparameter and test sets derive from a collection of 1,000 images that were seen by all subjects in the NSD. Training and validation sets originate from the remaining data. Our training stage prepares a decoding model to extract CLIP image embeddings $c_I$ from brain activity $\beta$, voxel-wise encoding models \cite{St-Yves2017} that map images $I$ to predicted brain activity $f(I) = \beta^\prime$, and a denoising autoencoder that maps measured brain activity $\beta$ to denoised brain activity $\hat{\beta}$. During inference, we average $\beta$ across three repetitions of the target image and decode to $\prime{c_I}$, which is held constant throughout the inference. We denoise the three fMRI data repetitions individually to produce $\hat{\beta}$, which we use to guide the search process. We use SD to generate a small library (n=250) conditioned on $\prime{c_I}$ and a latent $z_{t=0}$ sampled from a uniform Gaussian. We pass each image in the library, $I_j$, through the encoding models to obtain predicted brain activity $\beta^\prime_j$. We scored each image by computing Pearson correlation between the predicted activity pattern and $\hat{\beta}$. We select the images with the highest average correlation across the three repetitions, encode them as latent representations $z_{t=1}$, and generate a new library. With each iteration, we decrease the strength parameter of the diffusion model, which determines the relative level of guidance of the conditioning $c_I$ and $z_t$ variables during diffusion. For the examples shown here, we performed $10$ iterations, reducing the strength parameter along a cubic decay schedule from $1.0$ to $0.6$. We compare our search-based reconstruction procedure to a CLIP-only decoder in which we generate a single sample conditioned on $\prime{c_I}$, and a simple library search method in which the encoding model is used to select one image from a large (60K images) library of COCO images.

\begin{figure}[h]
\begin{center}
\includegraphics[scale=0.143]{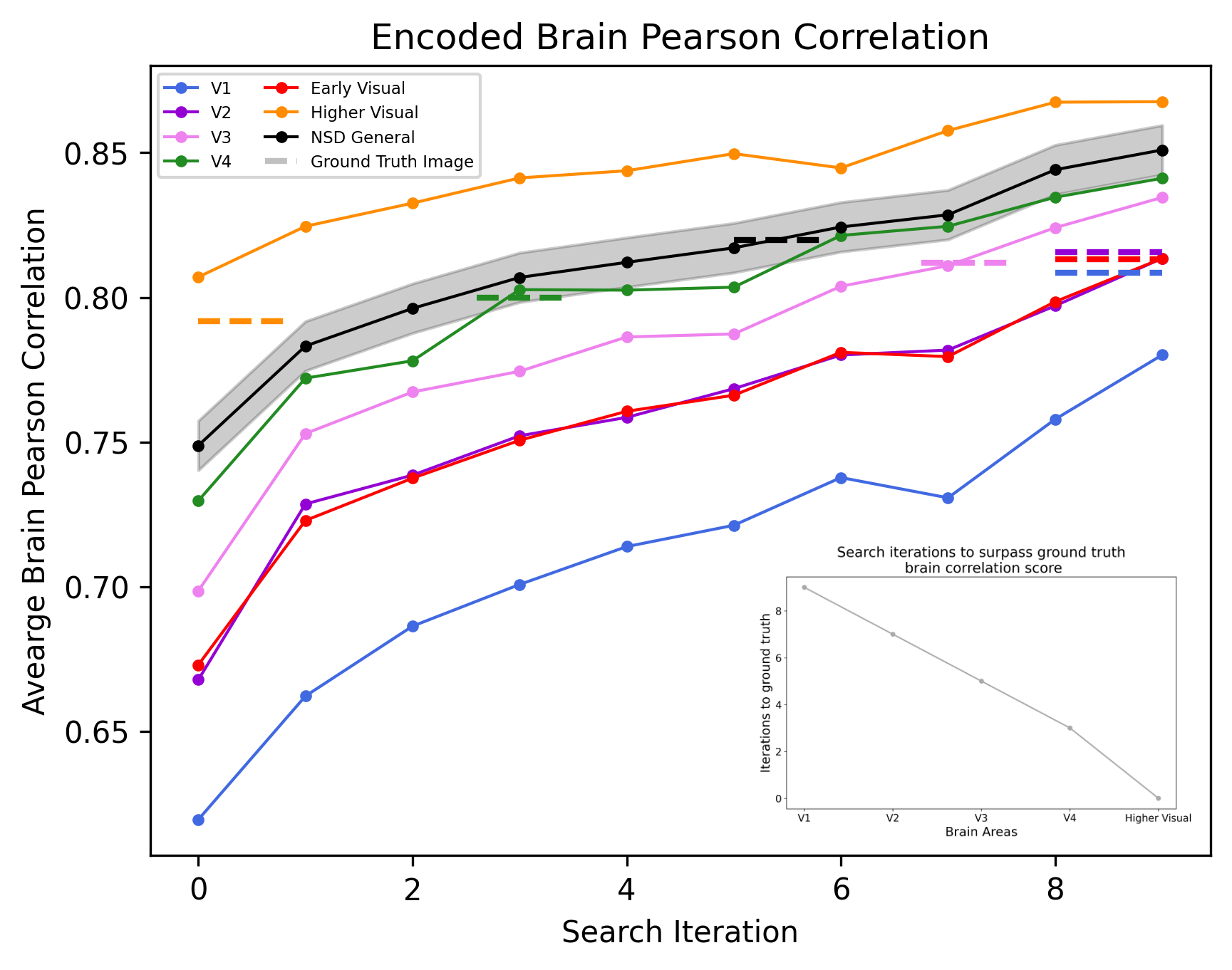}
\end{center}
\caption{Correlation of predicted and actual brain activity (``score") for the top image at each iteration for different ROIs (curves). Inset: the number of iterations for each ROI to cross score for ground truth image (dashed lines).} 
\label{figure:plot}
\end{figure}

\begin{table}[h]
\begin{center}
\resizebox{\columnwidth}{!}{%
\begin{tabular}{|llll|} 

\hline
Methods  &  PixCorr  & SSIM  & CLIP \\
\hline

Stochastic Search & $.215 \pm 0.041$ &  $.295 \pm 0.042$  &  $85.1 \pm 9.1$ \%  \\
CLIP Decoding     & $.067 \pm 0.038$ &  $.268 \pm 0.041$  &  $82.1 \pm 8.9$ \%  \\
Best COCO Image   & $.220 \pm 0.043$ &  $.277 \pm 0.067$  &  $82.0 \pm 9.0$ \%  \\
\hline
\end{tabular}%
}

\vskip 0.22in
\caption{Quantitative comparison of methods. PixCorr: pixel-wise correlation metric. SSIM: structural similarity index measure. CLIP: two-way identification experiment comparing CLIP from reconstruction to CLIP from ground truth target and a random reconstruction of the same type.}
\label{table:results}
\end{center}
\end{table}

\section{Results}
Preliminary experiments with our “semantics before structure” search converge on reconstructions that depict the right objects at roughly the right location and scale (Figure \ref{figure:results}) by refining low-level detail across search iterations (Figure \ref{figure:iterations}), outperforming simpler approaches on several metrics (Table \ref{table:results}).

We tracked the score for the best image at each iteration for different brain areas relative to the score for the ground truth target image. The number of search iterations required to achieve the score of the ground truth image decreased monotonically with progression from V1-V4 and into “high-level” visual cortex. This time-to-convergence provides a succinct measure of the level of representational invariance across visual cortex.

\section{Conclusion}
Our effort is one of several~\cite{Takagi2022.11.18.517004, lin2022mind,ozcelik2023braindiffuser,gu2023decoding, St-Yves_gan} that use recent developments in AI to achieve impressive reconstruction quality. Although further improvement is warranted, these efforts suggest that reconstructing seen images from human brain activity in controlled settings may soon be a solved problem. For better or worse, this brings promise of using brain decoding as a novel modality for communicating internal states incrementally closer to fulfillment.

\section{Acknowledgments}

Stability AI for providing code and pre-trained models; Professor Paul Schrater for guidance. This work was supported by NIH R01EY023384 (T.N.). Collection of the NSD dataset was supported by NSF IIS-1822683 and NSF IIS-1822929.

\bibliographystyle{apacite}

\setlength{\bibleftmargin}{.125in}
\setlength{\bibindent}{-\bibleftmargin}

\bibliography{ms}

\end{document}